# Extreme Kinematics of the 2017 September 10 Solar Eruption and the Spectral Characteristics of the Associated Energetic Particles

N. Gopalswamy[1], S. Yashiro[1,2], P. Mäkelä[1,2], H. Xie[1,2], S. Akiyama[1,2], and C. Monstein[3]

[1]NASA Goddard Space Flight Center, Greenbelt, MD 20771, USA

[2]The Catholic University of America, Washington DC 20064, USA

[3]Institute for Particle Physics and Astrophysics, ETH Zurich, Switzerland

ABSTRACT

We report on the 2017 September 10 ground level enhancement (GLE) event associated with a coronal mass ejection (CME) whose initial acceleration (~9.1 km s$^{-2}$) and initial speed (~4300 km s$^{-1}$) were among the highest observed in the SOHO era. The GLE event was of low intensity (~4.4% above background) and softer-than-average fluence spectrum. We suggest that poor connectivity (longitudinal and latitudinal) of the source to Earth compounded by the weaker ambient magnetic field contributed to these GLE properties. Events with similar high initial speed either lacked GLE association or had softer fluence spectra. The shock-formation height inferred from the metric type II burst was ~1.4 Rs, consistent with other GLE events. The shock height at solar particle release (SPR) was ~4.4±0.38 Rs, consistent with the parabolic relationship between the shock height at SPR and source longitude. At SPR, the eastern flank of the shock was observed in EUV projected on the disk near the longitudes magnetically connected to Earth: W60 to W45.

Key words: Sun: coronal mass ejections – Sun: radio radiation – Sun: particle emission

## 1. Introduction

Coronal mass ejections (CMEs) are energetic ejection of up to ~$10^{17}$ g of coronal material with speeds often exceeding 2000 km/s in to the heliosphere (e.g., Gopalswamy 2016 for a recent review). CMEs are thought to be responsible for large solar energetic particle (SEP) events via the shock acceleration mechanism as opposed to impulsive events accelerated in flares (e.g., Reames 2013). CMEs that accelerate rapidly and attain very high speeds close to the Sun tend to have hard fluence spectra (Bein et al. 2011; Gopalswamy et al. 2016; 2017). SEP events with ground level enhancement (GLE) thus have the hardest spectra because of the high CME speeds near the Sun and the shock formation height at ~1.5 solar radii (Rs) (Mewaldt et al. 2012; Gopalswamy et al. 2012; Nitta et al. 2012). On the other hand, slowly accelerating CMEs (typically originating from quiescent filament regions) form shocks at ~5.4 Rs from the Sun and have the softest spectra (Gopalswamy et al. 2015). Regular SEP events have a spectral hardness and shock formation height (1.7 Rs) that are intermediate between the above categories (Gopalswamy et al. 2017). The well-observed 2017 September 10 (Sep10) CME and the associated GLE event have provided an opportunity to test this hierarchical relationship and gain a better understanding of the CME-SEP association. The first appearance of the CME in the coronagraph field of view preceded the solar particle release (SPR) of GeV protons near the Sun, so we can determine the CME height at this time accurately. In particular, there was only one west-limb GLE event in cycle 23 (2001 April 18) in which the SPR occurred after the first appearance of the CME in the coronagraph FOV. Therefore, we have the opportunity to confirm the parabolic relationship (Reames 2009; Gopalswamy et al. 2013a) between the eruption longitude and the CME height at SPR by addition of a a new data point to fill the longitude gap in the range W80-W120 longitudes. This relationship is key to the understanding of particle acceleration by shocks and the magnetic connectivity to the observer, contrary to some claims that the particles are accelerated in the flare of the Sep10 event (Zhao et al. 2018). This event has already gained attention from the community because of the fast EUV wave (Seaton and Darnel 2018; Liu et al. 2018), SEP event detected both at Earth and Mars (Kurt et al. 2018; Guo et al. 2018), and extended solar gamma-ray emission (Share et al. 2017; Omodei et al. 2018; Gopalswamy et al. 2018). Schwadron et al. (2018) claimed that the SEP was one of the hardest in a decade, but it seems to be softer than seven large SEP events of cycle 24 (see Gopalswamy et al. 2016, Table 2). In this work, we report on the detailed evolution of the shock near the Sun and the characteristics of energetic particles that originate from the shock.

## 2. Observations

The Sep10 eruption occurred in NOAA AR 12673 located around S09W92 resulting in an ultra-fast CME (~3200 km s$^{-1}$) and an X8.3 soft X-ray flare starting, peaking, and ending at 15:35, 16:06, and 16:31 UT, respectively. EUV movies obtained by the Atmospheric Imaging Assembly (AIA, Lemen et al. 2012) on board the Solar Dynamics Observatory (SDO) indicate that the flare might have started slightly after ~15:40 UT when a small preceding flare peaked.

The eruption was associated with a spray-like eruptive prominence and a rapidly rising void in the interior of the white-light CME.

The CME and the leading shock were observed by the Sun Earth Connection Coronal and Heliospheric Investigation (SECCHI, Howard et al. 2008) on board the Ahead spacecraft (STA) of the Solar Terrestrial Relations Observatory (STEREO) mission. The eruption was observed by SECHHI's EUV Imager (EUVI), COR1 (inner coronagraph) and COR2 (outer coronagraph). In STA view, the eruption was ~50⁰ behind the east limb. The Large Angle and Spectrometric Coronagraph (LASCO, Brueckner et al. 1995) on board the Solar and Heliospheric Observatory (SOHO) observed the CME in its C2 and C3 telescopes. The CME first appeared at 16:00 UT in the LASCO/C2 FOV and STA/COR1 FOV.

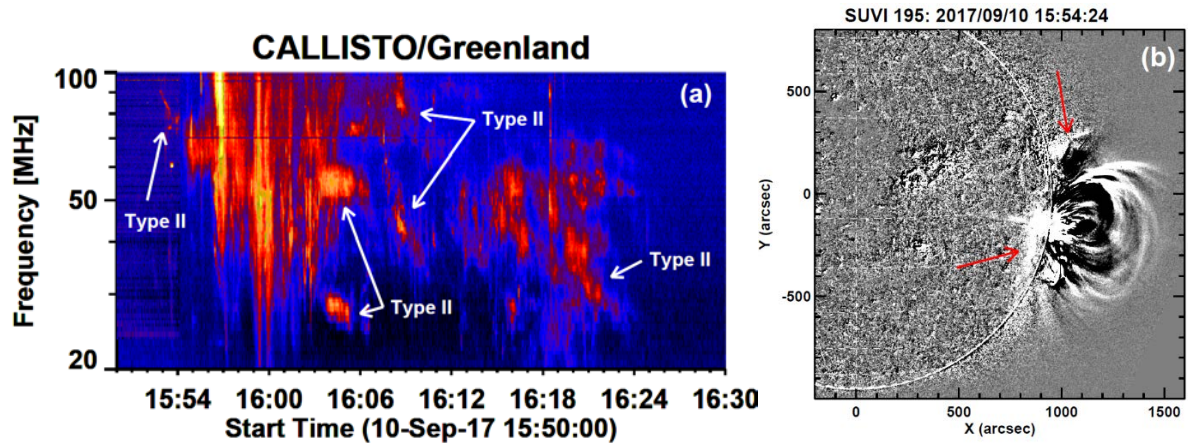

Figure 1. (a) Episode of metric type II emission from CALLISTO starting at 15:53 UT (90 MHz). The vertical features are type III bursts superposed on a type IV continuum. (b) An EUV composite image from GOES-16/SUVI (195 Å) showing the shock flanks above the limb and on the disk (pointed by the arrows).

The first shock signature was a metric type II radio burst at 15:53 UT observed by the Kangerlussuaq (Greenland) CALLISTO (Compound Astronomical Low Cost Low frequency Instrument for Spectroscopy and Transportable Observatory, Benz et al. 2009). Additional episodes of fundamental – harmonic structure at 16:04 and 16:08 UT and a broad-band structure before ending around 16:24 UT (see Fig. 1a) were observed. The type II emission was superimposed by type III bursts (the vertical streaks in Fig. 1a) and type IV emission (background continuum). The radio data have no spatial information, but the accurate onset time of the type II burst gives the shock-formation height (e.g., Gopalswamy et al. 2009a; 2012; 2013b). The EUV difference image at 195 Å from the NOAA's Solar Ultra Violet Imager (SUVI) in Fig.1b shows shock flanks to the north and east of the CME. This structure was not observed before type II onset. SUVI and AIA movies show these structures propagating as EUV waves across the disk (Seaton and Darnel 2018; Liu et al. 2018). The continued metric type II episodes at frequencies above 25 MHz until 16:24 indicates that the radio emission originated from shock flanks. Due to data gaps, the interplanetary type II burst was only observed starting September 11.

The eruption was associated with a large solar energetic particle (SEP) event with ground level enhancement (GLE). The >10 MeV proton intensity was ~1490 pfu, the fourth largest in cycle 24. None of the other three intense events had a GLE. The GLE was observed by several neutron monitors, including the one in Oulu (Usoskin et al. 2001), recording a 4.4%-increase above the background. Kurt et al. (2018) estimated a slightly larger increase (~6%) from other neutron monitors. The GLE was much weaker than the 2012 May 17 GLE (18.6%, Gopalswamy et al. 2013a) and only slightly stronger than the 2015 January 6 sub-GLE (2.5%, Thakur et al. 2014). The SEP event was observed in all GOES energy channels including the >700 MeV channel, which is indicative of GLEs (Gopalswamy et al. 2014a; Thakur et al. 2016).

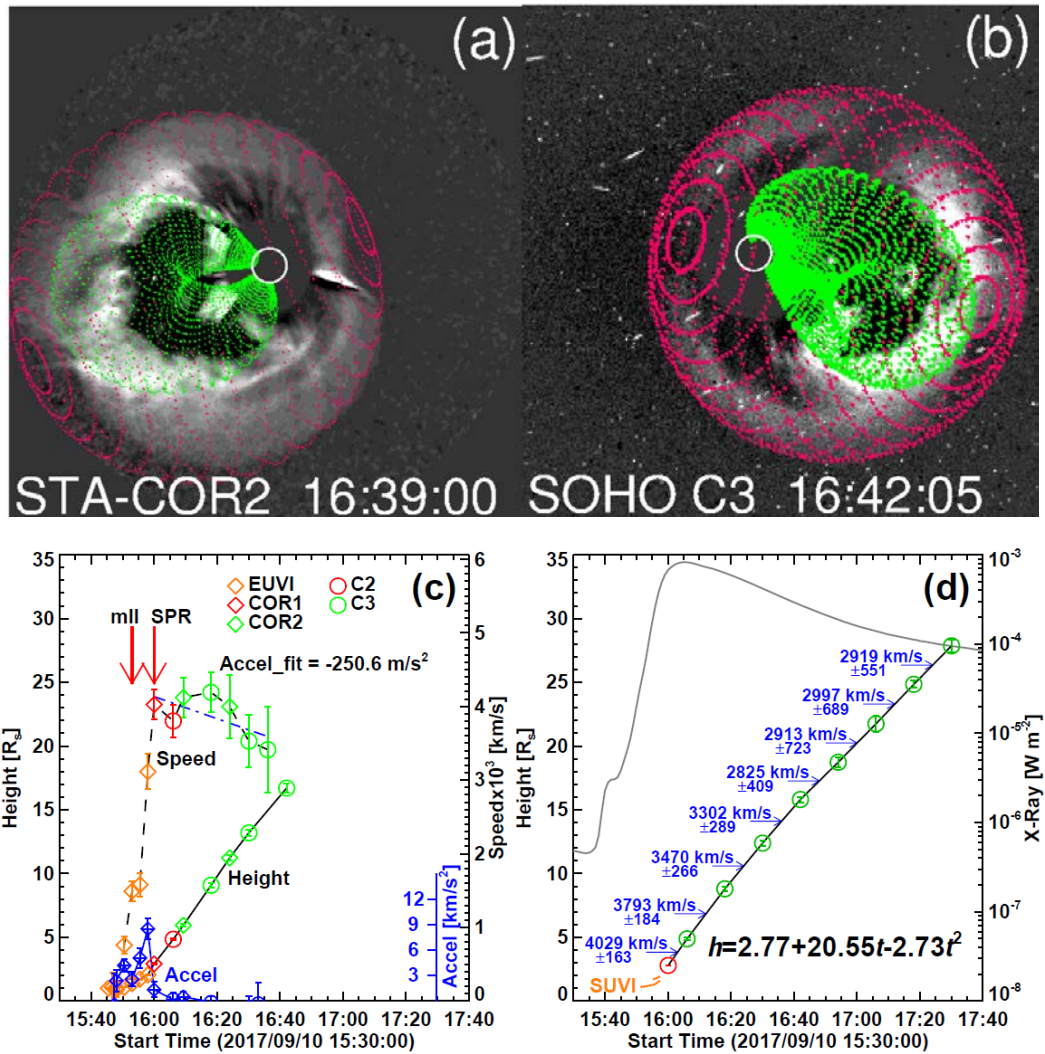

Figure 2. STEREO/COR2 (a) and LASCO/C3 (b) images with fitted flux rope (green) and shock (red) superposed. The direction, half width, and tilt angle of the flux rope are: S15W124, 50º, and 38º, respectively. (c) Height, speed, and acceleration of the shock from the GCS fit to SOHO and STEREO images. After peaking the speed declined at -250.6 m s$^{-2}$. Times of the metric type II burst (mII) and SPR are marked. By the time all the metric type II episodes ended (16:24 UT), the shock nose was beyond 10 Rs, suggesting that all metric type II emissions came from the shock flanks. (d) SOHO/LASCO height (h) - time (t) measurements. Speeds from successive h-t points are shown (at times pointed by arrows). The equation on the plot is from a quadratic fit to the h-t data. SUVI LE (upper) and void (lower) curves mark the start of the eruption.

## 3. Analysis and Results

We derive the CME/shock height-time history using LASCO sky-plane measurements and the Graduated Cylindrical Shell (GCS) model fit to the flux rope (Thernisien 2011) and spheroid to the shock (Olmedo et al. 2013). The CME leading edge (LE) denotes the shock. The solar particle release (SPR) time of GeV particles is determined assuming a Parker spiral length of 1.2 AU (Kahler 1994; Gopalswamy et al. 2012).

### 3.1 CME Kinematics

The coronal images obtained from STA and SOHO views are combined to fit a flux rope to the CME and a spheroid to the shock. After fixing the flux rope direction from simultaneous multiview observations, we extend the fitting to other images obtained earlier (e.g., EUVI/COR1) assuming there was no drastic change in the trajectory. Figure 2a,b shows the reconstruction with the fitted shock and flux rope superposed on STEREO and SOHO images. The shock height-time (h-t) history from the fit and LASCO sky-plane measurements are shown in Fig.2c,d. A straight-line fit to the LASCO h-t data gives an extremely high average speed: 3430±25 km s$^{-1}$, while a quadratic fit indicates a deceleration (0.29 kms$^{-2}$). The three-dimensional (3-D) speed attained >1000 km s$^{-1}$ in the first 10 minutes and ~3700 km s$^{-1}$ in the next 10 minutes (at 16:00 UT) (Fig. 2c). The acceleration attained a peak value of 9.1±1.6 km s$^{-2}$ at 15:58 UT, when the LE height was ~2.05 Rs with a speed of 3114±240 km s$^{-1}$. By the time the acceleration ceased (16:18 UT), the shock speed was ~4191±272 km s$^{-1}$. The peak acceleration is much higher than in previous reports: ~4.5 km s$^{-2}$ (SOHO, Zhang and Dere 2006) and ~6.78 km s$^{-2}$ (STEREO, Bein et al. 2011). Gopalswamy et al. (2016) reported on six LASCO CMEs with initial acceleration >5 km s$^{-2}$. Two of these were GLE events: 2003 October 28 (5.21 km s$^{-2}$) and 2005 January 17 (5.05 km s$^{-2}$). The other four were large SEP events with hard fluence spectra: 2011 August 4 (5.10 km s$^{-2}$), 2011 August 9 (6.93 km s$^{-2}$), 2012 July 7 (5.87 km s$^{-2}$) and 2012 July 8 (5.38 km s$^{-2}$). The lower LASCO cadence before 2010 August might have slightly affected the magnitude of the pre-2010 accelerations.

In summary, three speeds describe CME kinematics: (i) initial speed from the first two height-time data points (4029±163 km s$^{-1}$), (ii) maximum speed from the GCS fit (4191±272 km s$^{-1}$), and (iii) average speed within the LASCO FOV (3430±25 km s$^{-1}$). For comparison, the initial speeds of the two non-GLE events are 4038 km/s and 4416 km/s for the 2012 July 7 and 8 events, respectively. The initial speed is an indicator of the initial acceleration as explained in Gopalswamy et al. (2016).

### 3.2 The Energetic Particle Event

Figure 3a,b show SEP intensities in GOES-13 channels including the >700 MeV channel along with GLE intensity from the Oulu Neutron Monitor. The GLE intensity was at 10% of its peak at 16:08 UT, which we take as the onset time. The 5-min data in the GOES-13 highest energy channels (420 to 510 MeV, 510 to 700 MeV, and >700 MeV) indicate a slightly earlier onset (16:05 UT). We take 16:08 UT as the Earth arrival time of GeV particles. Guo et al. (2018) estimated a slightly later onset (16:15 UT). Assuming a Parker spiral length of 1.2 AU, the travel time of 1 GeV protons is ~11.3 min, so the SPR time is 16:05 UT, ~3 min behind the electromagnetic signals (white-light and EUV CME, type II burst). A 10% error in the Parker-spiral length will result in a 1.1-min error in the SPR time.

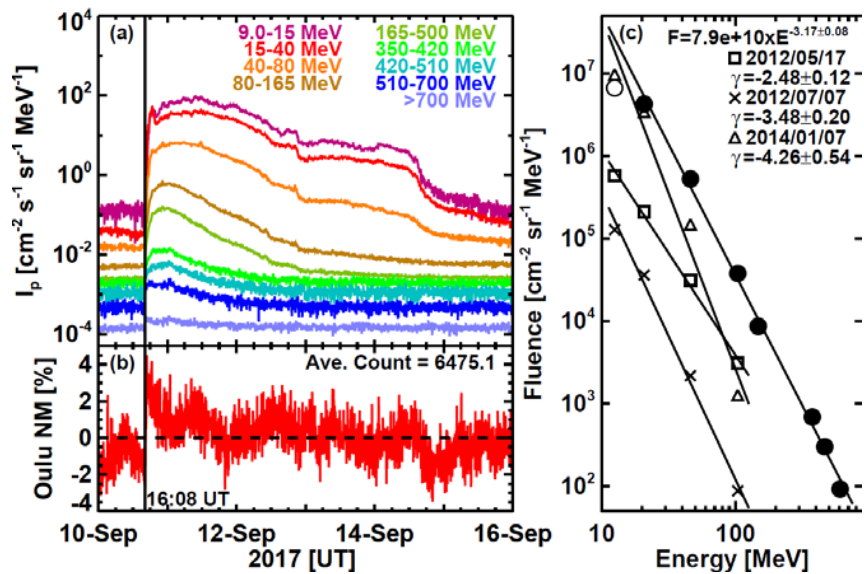

Figure 3. (a) Proton intensities in several GOES-12 energy channels. (b) GLE intensity (5-min averaged) above background count (from Oulu neutron monitor). The vertical line in (a) and (b) marks the GLE onset (16:08 UT). (c) Fluence spectra and spectral indices ($\gamma$) of the 2017 September 10 (filled circles), the 2012 May 17 (squares), 2012 July 7 (crosses) and 2014 January 7 (triangles) events. The open-circled data point was not included in the spectral fit because there seems to be a low-energy turnover at lower energies.

Figure 3c shows the fluence spectrum of the form: $F = AE^{-\gamma}$, where F is the fluence (cm$^{-2}$sr$^{-1}$MeV$^{-1}$), A ($7.9\times10^{10}$) is a constant, E is the particle energy and $\gamma$ is the spectral index (3.17±0.08). The spectrum is softer than that of the 2012 May 17 GLE ($\gamma$ = 2.48±0.12), but harder than those of the two non-GLE events: $\gamma$ = 3.48±0.20 (2012 July 7) and $\gamma$ = 4.26±0.54 (2014 January 7). The 2014 January 7 CME had an initial LE speed of ~4000 km s$^{-1}$, similar to that of the 2012 July 7 CME, but the initial acceleration was relatively low (~1.9 km s$^{-2}$).

### 3.3 The CME Leading-edge Height at SPR

The CME LE at SPR was at 4.4±0.38 Rs as inferred from the height-time history (Fig.2). The CME first appeared in the LASCO/C2 FOV at 16:00 UT at a height of 2.93 Rs and traveled for an additional 5 min to reach 4.4 Rs before SPR. The source longitude of GLEs ($\lambda$ in degrees) is related to the shock height at SPR ($h_s$ in Rs) by $h_s = 2.55 + [(\lambda-51)/35]^2$ (Gopalswamy et al. 2013a; Reames 2009). For the Sep10 event, $\lambda = 92°$, giving $h_s = 3.92$ Rs, which is only 11% smaller than the observed height (see Fig. 4b). Note that for well-connected events ($\lambda \sim 51°$), $h_s = 2.55$ Rs. Thus, the larger-than-average shock height at SPR is consistent with the parabolic relationship derived from cycle-23 GLEs. Furthermore, the Sep10 event provided an additional data point to reduce the longitude gap (86°-120°) in Fig.4b.

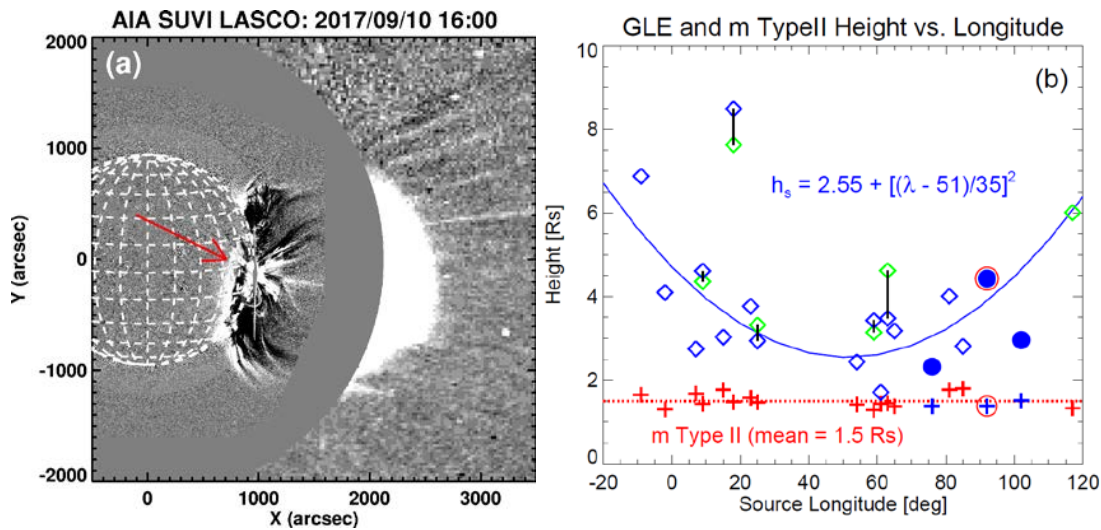

Figure 4. (a) SOHO/LASCO difference image at 16:00 UT, taken ~5 min before SPR, with simultaneous EUV (AIA, SUVI) difference images superposed. The eastward shock flank pointed by arrow) on the disk is between W60 and W45 at SPR. (b) CME/shock heights ($h_s$) at the time of SPR plotted against the source longitudes ($\lambda$) in GLE events from cycle 23 (diamonds) and cycle 24 (filled circles). The parabola is a fit to the cycle-23 data. The green diamonds represent cases in which SPR time was very close to the time of the LASCO image, so the heights were obtained without much extrapolation. The plus symbols denote the CME/shock height at the onset time of type II bursts (red – cycle 23; blue – cycle 24), with the horizontal dotted line denoting the mean (1.5 Rs). Shock heights of the 2017 September 10 event are shown circled.

At type II onset (15:53 UT, Fig.1a) the shock was at a height of 1.4 Rs. This is typical of GLE events in cycles 23 and 24 (plus symbols in Fig.4b), but the height at SPR varies because of the magnetic-connectivity requirement. After formation, the shock had to travel an additional 3 Rs during the 12 min before SPR. One part of the ~12 min is needed for the shock to accelerate particles to GeV energies; the other part is needed for the section of the shock releasing GeV particles to cross the Sun-Earth field line (Krucker et al. 1999; Rouillard et al. 2012). The latter is primarily responsible for the larger shock height at SPR. In fact, the EUV shock structure projected on the disk can be seen between W60 and W45 in Fig. 4a, around the SPR time. Zhao et al. (2018) claimed the type II onset to be at 16:03 UT, which led to the erroneous conclusion that the Sun-Earth magnetic field line was 1.7-AU long and the source of GeV particles was the associated flare.

### 3.4 Why the Soft Spectrum?

The average spectral index of GLE events is ~2.68 compared to 3.83 for non-GLE, western-hemispheric SEP events (Gopalswamy et al. 2016). The spectral index of the Sep10 GLE is 3.17, which indicates a softer-than-average spectrum for a GLE event. In spite of the extremely high initial speed and acceleration, the relatively soft spectrum and low GLE intensity are puzzling. In order to understand this, we compare our event with the 2001 April 18 GLE that originated from S17W120. This is the westernmost data point in Fig. 4b (SPR height ~6 Rs). Even though the longitudinal connectivity was poor, it had a hard spectrum ($\gamma = 2.59$). The B0 angle (the latitude of the solar disk center in heliographic coordinates) was -5º.4 for the 2001 April 18 event, which reduces the effective ecliptic distance to 11º.6. This is within the 13º ecliptic distance typical of GLE events (Gopalswamy et al. 2013a; 2014b; Gopalswamy and Mäkelä 2014). In the Sep10 event, the effective ecliptic distance was 16º.25 for a source at S09W92 and B0 = 7º.25. The distance is even larger (22º.3) when the flux rope latitude (S15) is considered. The flank at 16º.3 will have a lower speed, but high enough to accelerate particles to GeV energies because of the extreme nose speed. The large latitudinal extent implies a wider nose area, marginally connected to Earth resulting in the lower GLE intensity. On the other hand, lower energy particles are accelerated from all over the surface, resulting in a higher intensity and hence an overall softer spectrum. The weaker ambient field in cycle 24 compared to that in cycle 23 (Gopalswamy et al. 2014a; 2015) might also have contributed to the softer spectrum because of the reduced efficiency of shock acceleration.

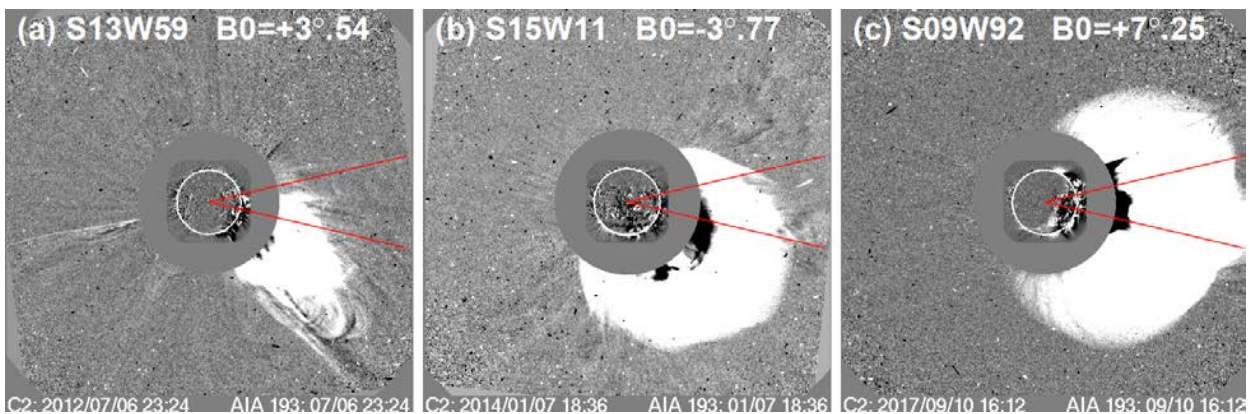

Figure 5. SOHO/LASCO CMEs in two non-GLE SEP events (a,b) whose initial speeds are similar to that of the 2017 September 10 CME (c). The red lines represent a cone of half angle of 13º based on the latitudes of cycle-23 GLEs. Note that the nose of the GLE CME is closer to the ecliptic than those of the other two.

We now compare our GLE event with the three non-GLE SEP events that had initial speeds exceeding 4000 km s$^{-1}$. Figure 5 compares the LASCO/C2 CME images. The 2012 July 7 CME erupted on July 6 from S13W59, but the nose was at position angle 233⁰, about 37⁰ from the equator. The B0 angle was +3.6⁰, so the ecliptic distance of the nose was 40⁰.6 indicating poor latitudinal connectivity. The flux-rope part of the CME was also unusually narrow. Therefore, only the remote northern flank, which is unlikely to accelerate GeV particles, was connected to Earth. The 2012 July 8 event was homologous to the July 7 event, so the same arguments apply. The 2014 January 7 disk-center CME (S15W11) had the nose at position angle of 231⁰. With B0 = - 3⁰.69, effective ecliptic distance was ~37⁰, as in the 2012 July 6 CME (Fig. 5). This CME was deflected to the south and west by coronal holes and the active region itself (Gopalswamy et al. 2014b; Möstl et al. 2015) such that only the remote northern flank was connected to Earth. GCS fits to these CMEs also indicate that the flux rope positions are different from the flare locations indicating deflection away from the ecliptic (Gopalswamy et al. 2014b). The flux rope directions were S32W62 (2012 July 7) and S15W29 (2014 January 7). In both events, coronal holes were present at the appropriate locations to account for the required deflection (Gopalswamy et al. 2009b). In comparison, the GLE-producing CME on 2014 January 6 had its nose right at the ecliptic indicating that the nose connectivity to Earth made the difference between this event and the one on January 7 (Gopalswamy et al. 2014b). The 2012 May 17 GLE had the opposite effect: a coronal hole located to the north of the eruption region (N11W76) deflected the CME toward the ecliptic such that the nose was at S07W76, very close to the ecliptic (see Gopalswamy et al. 2013a). As noted in Fig. 3c, the 2012 May 17 GLE had a hard spectrum even though the CME speed was only ~2000 km s$^{-1}$.

## 4. Summary and Conclusions

We studied the kinematics of the CME/shock associated with the Sep10 GLE to see how they are related to the SEP fluence spectrum. We also determined the shock formation height in the corona, taken as the CME leading-edge height at type II burst onset. Assuming a Parker spiral length between the Sun and Earth to be 1.2 AU, we computed the SPR time and the corresponding CME leading-edge height. The main findings of this paper can be summarized as follows:

1. The CME leading edge had an acceleration of 9.1±1.6 km s$^{-2}$, the highest ever observed in the STEREO era and probably ever since the CME phenomenon was discovered.

2. The initial speed, computed using the first two height-time data points was ~4029±163 km s$^{-1}$, also one of the highest in the SOHO era.

3. The CME leading edge was at a height of 1.4 Rs at the type II burst onset, consistent with all the GLE events of solar cycles 23 and 24. However, the CME leading edge at SPR (4.4±0.38 Rs) was larger-than-average primarily because of poor longitudinal connectivity.

4. The SPR time is consistent with the crossing of the Sun-Earth field line by the eastern flank of the CME-driven shock inferred from the EUV wave propagating across the disk.

5. The shock height at SPR is consistent with the parabolic relationship between shock height and source longitude derived from cycle-23 GLE events.

6. The GLE event had an intensity of ~4.4% above the background making it one of the small GLE events.

7. The SEP fluence spectrum had a power-law index of 3.17±0.08, which is larger than the average index of GLE events (2.68).

8. The low intensity and soft spectrum can be attributed primarily to the poor latitudinal and longitudinal connectivity.

**Acknowledgments.** STEREO is a mission in NASA's Solar Terrestrial Probes program. SOHO is a project of international collaboration between ESA and NASA. This work benefited from NASA's open data policy in using SDO data. We acknowledge the use of NOAA's GOES soft-Ray, SEP, and GOES-R Series Level 1b Solar Ultraviolet Imager (SUVI) data. The SUVI data are available at https://doi.org/10.7289/V5FT8J93. We thank K. Leer (Denmark) and P. Trottier (Greenland) for installing and operating the CALLISTO instrument, as part of the International Space Weather Initiative (ISWI) network. Oulu NM data are available at http://cosmicrays.oulu.fi. Work supported by NASA/LWS program. We thank the anonymous referee for helpful comments.